\documentclass[amsmath,prb,
twocolumn,citeautoscript]{revtex4-1}
\usepackage{bm}
\usepackage{graphicx}
\usepackage{amssymb}

\begin{document}
\title{Exact non-Hookean scaling of cylindrically bent elastic sheets\\ and the large-amplitude pendulum}
\author{Vyacheslavas Kashcheyevs}
\email[E-mail: ]{slava@latnet.lv}
\affiliation{Faculty of Physics
and Mathematics, University of Latvia, Zellu street 8, Riga LV-1002, Latvia}
\begin{abstract}
A sheet of elastic foil rolled into a cylinder and deformed between two parallel plates
acts as a non-Hookean spring if deformed normally to the axis. For large deformations the elastic
force shows an interesting inverse squares dependence on the interplate distance
{[\v{S}iber and Buljan, arXiv:1007.4699 (2010)]}.
The phenomenon has been used as a basis for an experimental problem at the 41st International
Physics Olympiad. 
We show that the corresponding variational problem
for the equilibrium energy of the deformed cylinder 
is equivalent to a minimum action description of
a simple gravitational pendulum with an amplitude of $90^{\circ}$. We use this analogy
to show that the power-law of the force is exact for distances less than a critical value. An analytical solution for the elastic force is found
and confirmed by measurements over a range of deformations covering both linear and non-Hookean behavior.
\end{abstract}
\maketitle

\section{Introduction}
\begin{figure}
  \begin{center}
  \includegraphics[width=6.5cm]{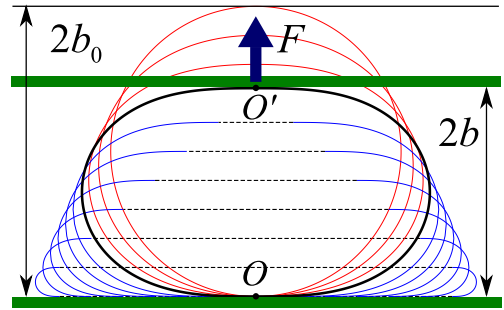}
  \end{center}
  \caption{\label{fig:fig1}
  A thin-walled elastic tube deformed between two parallel plates. The thick black contour
  line marks the profile at the critical value of $b_c/b_0=0.717770$,
  thinner lines above and below correspond to different $b/b_0$ between
  $0$ and $1$.}
\end{figure}

In a recent study \cite{Siber2010}, \v{S}iber and Buljan analyze
the following simple yet pedagogically rich problem from the theory of elasticity.
A thin  flat elastic sheet (e.g., a piece of plastic foil) is rolled into a cylinder (radius $b_0$) and placed between two
impenetrable plates which are parallel to each other and to the axis of the cylinder, see Fig.~\ref{fig:fig1}.
The distance $2 b$ between the plates is fixed externally. For  $ b <  b_0$ the foil acts as a spring exerting a force of magnitude $F(b)$ on
each of the plates. An interesting property for this kind of spring is the non-Hookean
power-law  scaling of the elastic force \cite{Siber2010}, $F \propto b^{-2}$, which holds
(as we show here, \emph{exactly}) for $b < b_c \simeq 0.7 b_0$.
Measuring $F(b)$ in this universal scaling regime has been proposed \cite{Siber2010}
as a method to determine bending rigidity for such objects as plastic foils, electrical connectors,
biological membranes and microtubules, possibly nanotubes and monolayer materials (e.g., graphene) etc.
A lab problem based on measuring of $F(b)$  for standard plastic transparency films has been recently
given to world's top secondary school physics students at the 41st International Physics Olympiad (Zagreb, Croatia, 2010).

The corresponding mathematical problem of constrained minimization of the foil's elastic energy may seem difficult and hardly illuminating.
Existing solution \cite{Siber2010} is based on analytical approximations and numerical finite element optimization while
a textbook approach \cite{LandauLifshitz7} relies on force equilibrium conditions for strongly bent elastic rods that are rarely
covered in standard physics curricula.

In this paper we formulate and solve the variational problem
using tangential angle parametrization of the profile shape.
This reveals equivalence to another conceptually rich physics system --- the large-amplitude pendulum:
elastic energy of the foil maps onto kinetic energy of the pendulum while the fixed inter-plate distance translates
into the cosine-shape potential energy.
Simple mechanical considerations allow us to deduce the exact inverse squares law for the elastic force at $b<b_c$.
Using the standard solution for the large-amplitude pendulum in terms of elliptic
functions, we find (a) the exact values for $b_c$ and related constants; (b)
a single transcendental equation that determines  $F(b)$ for $b> b_c$; and (c)
compact analytic form of the universal profile. Finally, the deduced functional dependence $F(b)$ is compared
to  measurements on a plastic film in the table-top setup used in Ref.~\onlinecite{Siber2010}. This show feasibility of
a quantitative demonstration of both the universal non-Hookean law for $b<b_c$ and the usual linear regime for $b\to b_0$.

\section{Formulation of the problem}
\begin{figure}
  \begin{center}
 \includegraphics[width=7.5cm]{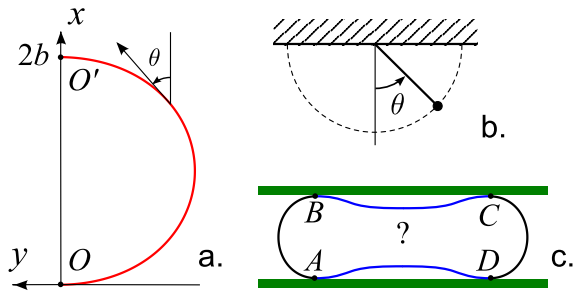}
  \end{center}
  \caption{\label{fig:fig15}
  (a) Coordinate system for profile energy calculation.
  (b) Equivalent pendulum problem with  $90^{\circ}$  maximal deviation.
  (c) A possible four contact-point profile.}
\end{figure}

Specific property of the deformation geometry in this problem  is ``cylindricity'': the Gaussian curvature vanishes at every point.
This eliminates non-uniform stretching/compression (\S 14 of Ref.~\onlinecite{LandauLifshitz7})
and leaves only the bending contribution to the total elastic energy  $W_{\text{el}}$ for thin sheets (thickness $d \ll b_0$).
The problem is essentially one-dimensional and the energy functional \eqref{eq:ElasticEnergy}
 is the same as for an elastic filament (Kirchhoff rod,  see \S 18 of Ref.~\onlinecite{LandauLifshitz7}):
\begin{align} \label{eq:ElasticEnergy}
  W_{\text{el}} = \frac{\kappa h}{2} \int \mathcal{K}^2 ds \, .
\end{align}
(This effects of gravity are ignored.)
Here $\kappa = (1-\nu^2)^{-1} E d^3 / 12$
is the bending rigidity, $E$ is the bulk Young modulus of the foil material,
 $\nu$ is the Poisson ratio,
$h$ is the length of the cylinder in the non-deformed direction (i.e.\ parallel to the axis),
and $\mathcal{K}$ is the planar curvature of the deformation profile in the plane normal to the axis ($x$-$y$ plane).
The shape of the profile is described using natural parametrization \cite{DiffGeomBook} $ \{ x(s) , y(s) \} $
in the coordinate system defined in Fig.~\ref{fig:fig15}a with origin at point $O$;
$s$ is the arc length measured counter-clockwise
form  $O$.
Integration in \eqref{eq:ElasticEnergy} is performed along the entire profile,
and the absence of stretching \cite{Note2} implies
that $\int ds  = 2 \pi b_0$ regardless of $b$.
The constraint imposed by the plates is expressed by
requesting that the vertical coordinate of point $O'$ is $2b$ (see Fig.~\ref{fig:fig15}a),
\begin{align} \label{eq:constraint}
  x(s\!=\!\pi b_0) = 2 b \, .
\end{align}

In the following we find the minimum of  $W_{\text{el}}$ subject to the constraint \eqref{eq:constraint}
and taking into account impenetrability of the plates. The elastic force is then obtained from $ F(b)= - d W_{\text{el}}/d (2 b)$.

\section{Analytic solution}
\subsection{Lagrangian formulation and analogy to the pendulum}
An intrinsic quantity characterizing the shape of a planar curve
is its tangential angle $\theta(s)$ defined as $\{ \dot{x} , \dot{y} \} =\{\cos \theta, \sin \theta \}$
(dot denotes derivative with respect to  $s$). Particular advantage of
employing $\theta(s)$ for the present problem is that the (signed) curvature
equals \cite{DiffGeomBook} to simply $\mathcal{K} = \dot{\theta}$.
Integrating $\dot{x}=\cos \theta$ over the right half
of the profile between the contact points $O$ and $O'$ gives the
expression of the  constraint \eqref{eq:constraint}
in integral form:
\begin{align} \label{eq:constraintintegral}
  \int_{0}^{\pi b_0} \! \cos \theta(s) \, d s = 2 b = \frac{2 b}{\pi b_0} \int_{0}^{\pi b_0} \! ds  \, .
\end{align}

At the points of contact between the foil and the plates, the tangent to the profile
must be parallel to the $y$-axis, hence the boundary conditions for $\theta(s)$,
\begin{align} \label{eq:bcs}
     \theta(s\!=\!0) =-\pi /2 \; ,  \, \theta(s\!=\!\pi b_0) =+\pi /2 \; .
\end{align}

A standard way of turning a constrained optimization problem to an unconstrained one
is the use of Lagrange  multipliers. In our problem both the target function $W_\text{el}$ and
the constraint \eqref{eq:constraintintegral} are expressed as integrals (functionals)
of the unknown function  $\theta(s)$.  Denoting the Lagrange multiplier for Eq.~\eqref{eq:constraintintegral}
by $\omega^2_0$ (for reasons that will become clear shortly)  and using left-right symmetry in Eq.~\eqref{eq:ElasticEnergy},
the variational problem becomes that of  unconstrained minimization
with respect to a function $\theta(s)$ and a number $\omega_0^2$ of the following
functional
\begin{align}
  \widetilde{W} = \frac{1}{2} \frac{W_{\text{el}}}{\kappa h} =
\int_0^{\pi b_0} \mathcal{L}\left [\theta(s), \dot{\theta}(s), \omega^2_0 \right ] \, ds
\end{align}
where
\begin{align} \label{eq:Lagrangian}
     \mathcal{L}= \frac{1}{2}\dot{\theta}^2+ \omega_0^2
      \cos \theta- \omega_0^2 \frac{2  b }{\pi b_0}  \, .
\end{align}

Now the analogy to the pendulum problem has become manifest.
Viewing $\widetilde{W}$ as dynamical action and $s$ as time,
one recognizes in Eq.~\eqref{eq:Lagrangian} the Lagrange function
of a rigid pendulum in uniform gravitational field with the angular frequency
of \emph{small} oscillations equal to $\omega_0$ (see schematic drawing in Fig.~\ref{fig:fig15}b).
In the pendulum problem,
$\theta$ is the angle of deviation from the stable equilibrium which must satisfy
the Newton's equation (Euler-Lagrange equation of the variational problem)
\footnote{For an alternative way to derive Eq.~(6), see problem 1 in \S 19
of Ref.~2},
\begin{align} \label{eq:Newtons}
    \ddot{\theta} + \omega_0^2 \sin \theta &= 0 \, .
\end{align}
In standard mechanics problems $\omega_0$ is usually given, and
the fulfillment of the boundary conditions \eqref{eq:bcs} is ensured by
choosing the appropriate initial velocity $\dot{\theta}(0) \equiv \mathcal{K}_0$
which is a function of $\omega_0$. In our case, both $\omega_0$ and $\mathcal{K}_0$
are not known \emph{a priori} and must be determined by satisfying both the boundary conditions \eqref{eq:bcs} and the constraint
\eqref{eq:constraintintegral}. The optimal value of $\omega_0^2$ is proportional \emph{the elastic force
itself} since, after ``integrating out'' the pendulum degree of freedom [i.e.\ substituting $\theta(s)$ in the Lagrange function
with the solution to the equation of motion \eqref{eq:Newtons}], $\widetilde{W}(\omega_0,b)$ satisfies
$\partial  \widetilde{W}/\partial \omega_0 = 0$ and thus
\begin{align}\label{eq:justforce}
  F = -\kappa h \frac{d \widetilde{W}(\omega_0,b)}{d b} = - \kappa h \frac{\partial \widetilde{W}}{\partial b} =2 \kappa h \omega_0^2\, .
\end{align}
We shall use this neat property of Lagrange multipliers for the calculation of $F(b)$.

Before tackling the problem of finding $\mathcal{K}_0$ and $\omega_0$ (which is mostly mathematical),
we develop physical arguments for the universal scaling behavior which constitute, in author's opinion,
the most important message of this article.

\subsection{Universal regime, $b \le b_c$}
Qualitatively, $\mathcal{K}_0(b)$ starts from the value of $\mathcal{K}_0(b_0)=1/b_0$ and
decreases as $b$ is decreased. At a ceratin critical (in the sense of separating two qualitatively different behaviors)
$b=b_c$ the contact
 curvature vanishes, $\mathcal{K}_0(b_c)=0$.
The corresponding pendulum problem becomes that of free oscillations with amplitude $\pi/2$
and frequency $\omega_0(b_c) = \omega_c$ (``critical pendulum''). Period of these oscillations
is well-known \cite{Belendez2010} [also derived below as a special case of Eq.~\eqref{eq:cond3}], $T_0 = 4 K(1/2)/\omega_c$,
where $K$ is complete elliptic integral of the first kind.
Since according to Eq.~\eqref{eq:bcs} it must be equal to $T_0 = 2 \pi b_0$, we get
\begin{align} \label{eq:omegac}
   \omega_c & = (\zeta_0 \, b_0)^{-1} \, \text{ with } \zeta_0 \equiv \frac{\Gamma^2(3/4)}{\sqrt{\pi}}= 0.847213 \ldots  \, .
\end{align}

When $b$ is decreased further below  $b_c$, an extra condition not accounted for by the Lagrangian formulation \eqref{eq:Lagrangian}
becomes relevant: the plates do not allow the foil to bend outwards,
thus $\mathcal{K}_0$ remains zero
also for $b < b_c$. In order to accommodate the imposed small values of $b$
without violating the rigid plates, a finite part of the foil in the vicinity of $s=0$ and $s= \pi b_0$
must remain flat, $\mathcal{K}(s)=0$ (these parts are marked by horizontal dashed lines in the profiles
shown in Fig.~\ref{fig:fig1}). Concurrently, the sections of the profile
that do bend obey Newton's equation \eqref{eq:Newtons}, although with some  $\omega_0(b) > \omega_c$.
Assuming continuity in $\mathcal{K}(s)$, we conclude that the deformed part of the profile for $b<b_c$ must start with $\mathcal{K}_0=0$
(same as for $b=b_c$) but cover a length shorter than the full length $2 \pi b_0$ of the foil cross-section. This shortening of the deformed part
can only
be accommodated  by a faster pendulum since the frequency $\omega_0(b)$ of the latter
remains the only adjustable (i.e.\ $b$-dependent) parameter in the problem for $b<b_c$. (The other
parameter, $\mathcal{K}_0$, is pinned to zero by the presence of a finite flat part.)
But, according to  Eq.~\eqref{eq:Newtons}, changing  $\omega_0$ results in a \emph{uniform} rescaling of the arc length parameter $s$,
therefore the bent parts of the profile at $b<b_c$ (marked by
curved lines between the plates in Fig.~\ref{fig:fig1})
must be
\emph{geometrically similar}
to the corresponding halves of the critical profile at $b=b_c$
(marked with a thick contour in Fig.~\ref{fig:fig1}).

Having established that the \emph{critical} profile shape is also \emph{universal} (in the sense of being applicable to any $b<b_c$),
we can determine the scaling law for the force using simple dimensional considerations.
Scaling of $b$ and $b_0$ with $b/b_0$ fixed does not change the overall shape
of the profile  --- the tangential angle $\theta$ remains the same function of $s/b_0$.
Therefore the elastic energy \eqref{eq:ElasticEnergy} can be written as \cite{Siber2010}
$W_{\text{el}}(b) = \kappa h \, \mathcal{U}(b/b_0)/b_0$ where $\mathcal{U}$ is
a dimensionless function of $b/b_0$. For $b<b_c$
the flat parts of the profile do not contribute to the energy.
The curved  parts are similar to the critical profile but $b/b_c$ times smaller, thus
their energy must have a fixed  $\mathcal{U}(b/b_0) \to \mathcal{U}(b_c/b_0)$ and rescaled
 $b_0 \to b_0 \, b/b_c$. This argument proves exact scaling for the energy,
$W_{\text{el}} \propto b^{-1}$, and, consequently,  the force,
$F \propto b^{-2}$ for $b<b_c$. The prefactor in the scaling law can be described as follows:
\begin{align} \label{eq:scaleuniversal}
  F(b) = F_c \,  \frac{b_c^2}{b^2} =2 \kappa h \frac{ (b_c/b_0)^2}{\zeta_0^2 }  b^{-2} \text{ for } b<b_c \, .
\end{align}
where $F_c \equiv F (b\!=\!b_c)$ and  Eqs.~\eqref{eq:justforce} and \eqref{eq:omegac} have been used.
The phenomenological ``stadium profile'' model, considered as a variational \emph{ansatz}
in Ref.~\onlinecite{Siber2010}, assumes circle as a universal profile and  predicts $F(b) = (\pi/2) \kappa h/b^{-2}$, similar
to Eq.~\eqref{eq:scaleuniversal}.
The exact critical shape is, however, more efficient than a circle in minimizing the bending energy.
We can calculate the numerical prefactor in Eq.~\eqref{eq:scaleuniversal} exactly by
determining $b_c$. To this end, we proceed to the integration of the pendulum's equation of motion.

The first integral of Newton's equation is the energy conservation law,
$ \dot{\theta}^2/2 - \omega_0^2 \cos \theta = \text{const}$.
Taking into account $\cos \theta(s\!=\!0) =0$ gives
  \begin{align} \label{eq:energyConservation}
    \frac{1}{2} \dot{\theta}^2   & = \frac{1}{2} {\mathcal{K}_0}^2 + \omega_0^2 \cos \theta \, .
  \end{align}
  Integrating  both sides of Eq.~\eqref{eq:energyConservation} with respect to $s$ and with respect to $\theta$, then
  using Eq.~\eqref{eq:constraintintegral} gives
  \begin{align}
    \widetilde{W} & =  \frac {1}{2}\int\limits_0^{\pi b_0} \dot{\theta}^2 \, ds  = \frac{\pi b_0 \mathcal{K}_0^2}{2} + 2 \omega_0^2 b
    \label{eq:cond1} \, , \\
    \widetilde{W}
     &= \frac{1}{2} \int_{-\pi/2}^{+\pi/2} \dot{\theta} \, d \theta
     = \omega_0 I_{0}(\mathcal{K}_0/\omega_0) \label{eq:cond2} \, ,
  \end{align}
  where \begin{align}
    I_{n}(\alpha) \equiv (1/2) \int_{-\pi/2}^{+\pi/2} (\alpha^2 + 2 \cos \theta)^{1/2-n} d\theta.
  \end{align}
  
   Now $b_c$ can be determined from Eqs.~\eqref{eq:cond1} and \eqref{eq:cond2}:
  substituting $b_c$, $\omega_c$, and $0$ for $b$, $\omega_0$, and ${\mathcal{K}_0}$
  respectively, and using an identity $\zeta_0 = I_0(0)/2$, one gets
  \begin{align} \label{eq:mybc}
      b_c = I_0(0)/(2 \omega_c) = \zeta_0^2 \, b_0 = 0.717770 \ldots  \times b_0 \, .
  \end{align}
  The results \eqref{eq:scaleuniversal} and \eqref{eq:mybc} confirm phenomenological Eq.~(17)
  of Ref.~\onlinecite{Siber2010} (with numerical factor $0.912$ there corrected to
   $4\zeta_0^2/\pi = 0.91389$).

\subsection{Small deformations, $b_0\! >\! b\! >\! b_c$}
   In this regime $\mathcal{K}_0>0$
   and the relation between the period and the boundary conditions
   becomes more complicated.
   Expressing $\omega_0 \, d s$ via $d \theta$ from
   Eq.~\eqref{eq:energyConservation} and integrating over from $s=0$ to $s=\pi b_0$,
   we obtain the required additional equation,
  \begin{align} \label{eq:cond3}
    \pi b_0 \omega_0 & = 2 I_{1}(\mathcal{K}_0/\omega_0) \, .
  \end{align}
  This equation will provide the answer for the force, thanks to Eq.~\eqref{eq:justforce}.
  The remaining unknown is the parameter $\alpha \equiv \mathcal{K}_0/\omega_0$,
  for which a single  transcendental equation  is easily derived by combining Eqs.~\eqref{eq:cond1}, \eqref{eq:cond2}, and \eqref{eq:cond3},
  \begin{align} \label{eq:alphaeq}
    \frac{b}{b_0} & = \frac{\pi}{4} \left ( \frac{I_{0}(\alpha)}{I_{1}(\alpha)} - \alpha^2 \right ) \, .
  \end{align}

\begin{figure}
\begin{center}
  \includegraphics[width=7cm]{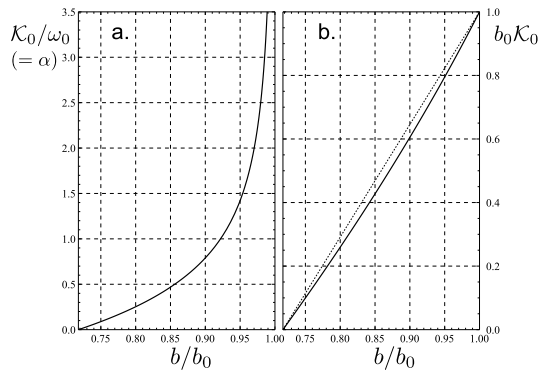}
  \end{center}
   \caption{\label{fig:fig3}
  (a) Graphical solution of Eq.~\eqref{eq:alphaeq}.
  (b) The curvature at the contact point $\mathcal{K}_0$ as a function of $b/b_0$. The dashed straight line
  $(b-b_c)/(b_0-b_c)$ is plotted for visual guidance.}
\end{figure}

  The functions $I_n$, defined as integrals and appearing in Eqs.~\eqref{eq:cond2}, \eqref{eq:cond3} and \eqref{eq:alphaeq},
  can be reduced to standard elliptic integrals \cite{AbramowitzStegunElliptic},
  $I_0(\alpha) = (4/k) \mathcal{E}(\pi/4,k^2)$ and  $I_1(\alpha) =  k \mathcal{F}(\pi/4,k^2)$,   where
  $\mathcal{E}$ and $\mathcal{F}$ are incomplete elliptic integrals of the first and the second kind, respectively, and
  $k\equiv 2/\sqrt{2+\alpha^2}$ is the elliptic modulus. $I_1(0) = K(1/2)=\pi / (2 \zeta_0)$ in consistence with Eq.~\eqref{eq:omegac}.
  We were unable, however, to solve the transcendental Eq.~\eqref{eq:alphaeq} analytically and present a graphical solution instead, see
  Fig.~\ref{fig:fig3}a.

As $b/b_0$ changes from $b_c/b_0$ to $1$, the only root of
Eq.~\eqref{eq:alphaeq} goes from $0$ to $\infty$.  Divergence of $\alpha = \mathcal{K}_0/\omega_0$ as
$b \to b_0$ is consistent with $\omega_0^2 \propto F \to 0$. The curvature at the contact point
$\mathcal{K}_0= \alpha \omega_0$ is plotted separately in Fig.~\ref{fig:fig3}b;
$\mathcal{K}_0$ decreases from $1/b_0$ to $0$ as
$b$ goes from $b_0$ to $b_c$, in accordance with our previous qualitative discussion.

Summing up the solution for $b>b_c$, the exact elastic energy and the force are
\begin{align}
   W_\text{el}& =\frac{ 4 \kappa h }{\pi b_0} I_0(\alpha) I_1(\alpha)
   \, , \label{eq:largeenergy}\\
  \label{eq:largeForce}
  F & = \frac{2 \kappa h }{b_0^2}  \left ( \frac{2 I_1(\alpha) }{\pi} \right )^2
  \, ,
\end{align}
where $\alpha(b/b_0)$ is the root of Eq.~\eqref{eq:alphaeq}, see Fig.~\ref{fig:fig3}a.

\subsection{Stability of the universal solution and bending-induced tension}
After formulating an explicit solution for $b>b_c$,  Eq.~\eqref{eq:largeenergy},
we can justify the universal scaling for $b< b_c$ more rigorously.
It has been suggested \cite{Siber2010} that at $b \lesssim 0.3 b_0$ profiles with a slight
curvature in the horizontal part may provide a better solution than a ``stadium'' with completely flat
segments. Such a four-contact-point  alternative solution is sketched in Fig.~\ref{fig:fig15}c
(the concave segments $B C$ and $A D$ are exaggerated).
The following argument shows, however, that any four-contact-point profile will
have larger bending energy than the optimal profile with straight $B C $ and $A D$.

Let us fix the positions of the contact points on the profile, e.g., by gluing the foil to the
plates at  $A$, $B$, $C$, and $D$,
The curvature $\mathcal{K}_A$
must be the same at all of these four points due to symmetry; furthermore,
$\mathcal{K}_A$ is evidently positive.
Thus the strongly curved parts $AB$ and $CD$
must conform to our solution for $\mathcal{K}_0 =\mathcal{K}_A >0$ with
a smaller $b_0$, adjusted to their respective shorter lengths.
Consider now increasing slightly the lengths of  $AB$ and $CD$
while making $BC$ and $AD$ shorter so that the total circumference
of the profile remains the same. In this new configuration
the energy of the segments
$BC$ and $AD$ will evidently be reduced by shortening.
However,
the corresponding lengthening  of the segments  $AB$ and $CD$ shall \emph{also reduce} their energy:
for fixed $b$, longer (higher $b_0$) two-point profiles have \emph{lower} minimal total energy.
This can be shown explicitly using Eqs.~\eqref{eq:justforce}, \eqref{eq:cond3}, \eqref{eq:largeenergy} and \eqref{eq:largeForce}:
\begin{align}\label{eq:Fvalue}
 F_T \equiv \left . \frac{d W_\text{el}}{d (2 \pi b_0) } \right \rvert_{b=\text{const}} = - \frac{\kappa h}{2} \mathcal{K}_0^2 \leq 0 \, .
\end{align}
Thus one can always lower the bending energy of a four-point profile
like the one shown in Fig.~\ref{fig:fig15}c down to the point where
$BC$ and $AD$ become completely flat and can no longer be shortened.
This minimal energy limit is precisely our zero-contact-curvature
universal stadium profile.

It is instructive to note that $F_T/h$ is, by definition, the surface tension energy of the deformed foil.
Uniform tension may be induced in a cylindrically deformed sheet by bending
even if the stretching deformation is negligible \cite{LandauLifshitz7}.
The value \eqref{eq:Fvalue} for  $F_T$ and its independence of $s$ can be confirmed using a formal method of
local Lagrange multiplies \cite{Cebers2003} which takes local tension into account explicitly.

Recalling that our sheet is assumed to be made of uniform elastic material
with bulk Young modulus $E$, we can now estimate the neglected stretching deformation.
Bending-induced negative surface tension is balanced by uniform linear stretching.
By using Hooke's law, we can express the increment in the profile circumference $ 2 \pi \Delta b_0$
due to stretching as
$\Delta b_0 = b_0 F_T/(E d h)$.
The corresponding contribution of stretching
$\Delta W_{\text{stretch}} \sim E h d^5/b_0^3 $ to the total energy $W \sim E h d^3/b_0$
is negligible if $d^2 \ll b_0^2$, thus confirming our approximation of an unstretchable foil.
The universal profile is tension free, $F_T(b \le b_c)=0$.

\subsection{Critical shape}
Before discussing our final results for the force, Eqs.~\eqref{eq:scaleuniversal} and \eqref{eq:largeForce}, we
briefly comment on the shape of the critical profile. Using the explicit solution of the pendulum
problem [obtained, e.g., by integrating Eq.~\eqref{eq:energyConservation}, see Ref.~\onlinecite{Belendez2010} for a pedagogical presentation],
and using properties of the elliptic functions \cite{AbramowitzStegunElliptic}, the shape of the profile at $b=b_c$, in the units corresponding to $b_0 =1/(\zeta_0 \sqrt{2})$,
is given parametrically for $\theta \in [-\pi/2, \pi/2 ]$  by
\begin{align} \label{eq:Uprofile}
\begin{cases}
  x(\theta) & =\zeta_0 /\sqrt{2} + \mathcal{E} \left ( \theta/2, \sqrt{2} \right ) \,  , \\
  y(\theta) & = \pm \sqrt{ \cos \theta}  \,  ,
  \end{cases}
\end{align}
with the arc length $s(\theta) = \pi b_0/2 +\mathcal{F}(\theta/2,\sqrt{2})$.
This shape is shown in Fig.~\ref{fig:fig1} by the thick contour touching the plates.
Profile shapes for  $b>b_c$ are obtained by integrating Eq.~\eqref{eq:Newtons} numerically
with the initial condition obtained from Eqs.~\eqref{eq:cond3} and \eqref{eq:alphaeq} and shown
by thin
lines reaching above the upper plate in Fig.~\ref{fig:fig1}.

\section{Discussion and comparison to experiment}
\begin{figure}
  \begin{center}
  \hspace{0.3cm}\includegraphics[width=7.2cm]{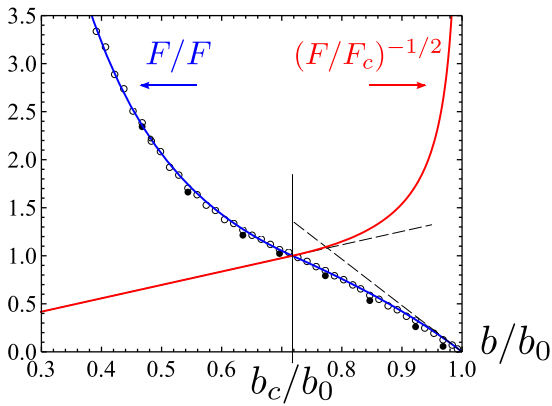}
  \end{center}
   \caption{\label{fig:fig2}
  {Elastic force $F$ as a function of the half-height $b$, scaled by the critical force $F_c$ and
  the undeformed cylinder radius $b_0$, in linear and inverse square root representations.}
  Circles: experimental data scaled by a single fitting parameter, $F_c$. The spring was first gradually loaded ($\circ$) up to
  $b_\text{min}=0.4 \, b_0$  and then unloaded ($\bullet$).}
\end{figure}
The final results for the force are shown in Fig.~\ref{fig:fig2}. We use $F_c = 2 \kappa h/ (\zeta_0 b_0)^2$
to scale the force into a dimensionless form, $F(b)/F_c$, which is a single universal function of $b/b_0$.
This function is shown by continuous lines in Fig.~\ref{fig:fig2} both in terms of $F(b)/F_c$
and
$(F(b)/F_c)^{-1/2}$ to reveal the region of power-law scaling.
The point at $b=b_c$ is an inflection point of $F(b)$ (with a jump in the third derivative).

It is instructive to verify that the rolled foil behaves as a Hookean spring when close to cylindrical shape (i.e.\ as $b \to b_0$).
Using the definition of  $I_n(\alpha)$ to get the large $\alpha$ expansion, we obtain $\alpha$ from Eq.~\eqref{eq:alphaeq},
$b_0-b \sim (\pi/4- 2/\pi) \alpha^{-2} b_0$,  and $F(b)$ from Eq.~\eqref{eq:largeForce} as follows
\begin{align}
  F(b) = \frac{2 \kappa h }{b_0^2}  \frac{b_0-b}{b_0} \frac{ 4 \pi}{\pi^2-8} \, \text{ for } b \to b_0 \, .
\end{align}
This linear behavior is marked by a dashed line in lower right corner of Fig.~\ref{fig:fig2}.

Measurements of $F(b)$ have been performed using the setup and one of the samples
(blue plastic binding covers, Set 1) described in Ref.~\onlinecite{Siber2010}. The results are shown
in Fig.~\ref{fig:fig2}  by circles on a \emph{linear scale}  (cf.\ the logarithmic scale used
in Fig.~5 of Ref.~\onlinecite{Siber2010}).
The spring was gradually loaded from $b=b_0$ down to $b_{\text{min}}=0.4 b_0$ (open circles) and
then unloaded back by  increasing $b$ up to zero force (filled circles).
Note the hysteresis due to inelastic deformations.
The corresponding energy
losses (hysteresis loop area) are $4 \%$ of $W_{\text{el}}(b_{\text{min}})$.
The critical force $F_c=116 \text{ N}$ with relative error of $3 \%$
(estimated from the residuals between the open data points and the
analytical fit) corresponds to the bending rigidity of $\kappa= (1.48 \pm 0.04) \text{ mJ}$
in reasonable agreement with Ref.~\onlinecite{Siber2010}.


\acknowledgments
The author is thankful to Paul Stanley and Eli Raz for inspiring discussions of the problem.


\end{document}